# Set Covering with Ordered Replacement
## Additive and Multiplicative Gaps

Friedrich Eisenbrand[a], Naonori Kakimura[b], Thomas Rothvoß[a], Laura Sanità[a]

[a] EPFL, Lausanne, Switzerland
[b] University of Tokyo, Japan

**Abstract**

We consider set covering problems where the underlying set system satisfies a particular replacement property w.r.t. a given partial order on the elements: Whenever a set is in the set system then a set stemming from it via the replacement of an element by a smaller element is also in the set system.

Many variants of BIN PACKING that have appeared in the literature are such set covering problems with ordered replacement. We provide a rigorous account on the additive and multiplicative integrality gap and approximability of set covering with replacement. In particular we provide a polylogarithmic upper bound on the additive integrality gap that also yields a polynomial time additive approximation algorithm if the linear programming relaxation can be efficiently solved.

We furthermore present an extensive list of covering problems that fall into our framework and consequently have polylogarithmic additive gaps as well.

## 1 Introduction

SET COVER is a prominent combinatorial optimization problem that is very well understood from the viewpoint of multiplicative approximation. There exists a polynomial time factor $O(\log n)$ approximation for SET COVER [2] and a corresponding hardness result [10]. Also the (multiplicative) integrality gap of the standard linear programming relaxation for SET COVER is known to be $\Theta(\log n)$ [17].

Let $\mathcal{S}$ be a family of subsets of $[n] = \{1, \ldots, n\}$, $w : \mathcal{S} \to \mathbb{R}_+$ be a cost function and let $\chi(S) \in \{0,1\}^n$ denote characteristic vector of a set $S \in \mathcal{S}$. The SET COVER integer program

$$\min \Big\{ \sum_{S \in \mathcal{S}} w(S) x_S \mid \sum_{S \in \mathcal{S}} x_S \cdot \chi(S) \geq \mathbf{1}, x \geq \mathbf{0}, x \text{ integral} \Big\} \tag{1}$$

and its linear programming relaxation is also in the focus of this paper. However, we are interested in the *additive gap* of a certain class of set covering problems. This additive gap is the *difference* between the optimum value of the integer program (1) and its linear programming relaxation. While there exists an extensive amount of literature on the (multiplicative) gap and (multiplicative) approximation algorithms, the additive gap and algorithms to construct integer solutions that are within the corresponding additive range have received less attention.

Why is it interesting to study the additive integrality gap of set covering problems? Suppose, for example that we know of a certain class of set covering problems that the additive gap is polylogarithmic, $\log n$ say. If we then, at the same time, know that the optimum solution is at least $\sqrt{n}$, then the linear programming relaxation of (1) asymptotically approaches the optimum solution of the integer program yielding a $(1 + \log n/\sqrt{n})$-factor approximation algorithm if an integer solution respecting the gap can be efficiently computed.

Two prominent covering problems whose additive gap has been studied are MULTI-EDGE COLORING [15, 19] and BIN PACKING.[1] For BIN PACKING, Karmarkar and Karp [16] showed that the additive gap is bounded by $O(\log^2 n)$ and they also provide a polynomial time algorithm that constructs a solution within this range. There is an extensive amount of literature on variants of BIN PACKING (see e.g. [8, 7, 6, 9, 8, 4, 22, 1]). The question whether the SET COVER linear programming relaxations of such variants also exhibit small additive gaps is in the focus of our paper.

---

[1] Even though coined bin "packing", it is a covering problem.

It is easy to see that the additive gap of general SET COVER is $\Theta(n)$. For example, the VERTEX COVER problem on a disjoint union of triangles exhibits this additive gap. What makes BIN PACKING so special that polylogarithmic additive gaps can be shown to hold? It turns out that it is essentially the fact that in a feasible packing of a bin, we can replace any item by a smaller item and still remain feasible. In the setting of SET COVER this is reflected by the following. There is a partial order $\preceq$ of the elements that we term *replacement order*. The order is *respected* by $\mathcal{S}$ if

$$S \in \mathcal{S}, i \in S, j \notin S, j \preceq i \Rightarrow ((S\setminus\{i\}) \cup \{j\}) \in \mathcal{S}$$

We will also consider costs $w(S)$ of sets in the family $\mathcal{S}$. These costs are normalized in the sense that $w(S) \in [0,1]$ for each $S \in \mathcal{S}$. The costs *respect* the replacement order if $w(S) \geq w(S')$ whenever $S'$ is obtained from $S$ by replacing one element $i \in S$ with an element $j \preceq i$ and if $w(S') \leq w(S)$ for any $S' \subseteq S$. Given a family $\mathcal{S}$ and costs respecting the replacement order $\preceq$, the SET COVER WITH ORDERED REPLACEMENT problem is to solve the integer program (1). We denote the optimum value of (1) and its relaxation by $OPT(\mathcal{S})$ and $OPT_f(\mathcal{S})$, respectively. The additive gap of $\mathcal{S}$ is thus $OPT(\mathcal{S}) - OPT_f(\mathcal{S})$.

**Contributions**

We provide a rigorous account on additive and multiplicative integrality gaps and approximability of SET COVER WITH ORDERED REPLACEMENT if $\preceq$ is a total order. Our main results are as follows.

- We show that the additive gap is bounded by $O(\log^3 n)$. This is achieved by the use of suitable *pseudo sizes* and grouping. The pseudo sizes are responsible for the additional $\log n$ factor compared BIN PACKING. If *natural sizes* are available, our bound matches the $O(\log^2 n)$ bound for BIN PACKING. The grouping technique itself is not novel albeit appears to be simpler as the one in [16].

- We provide a $\Omega(\log n)$ lower bound on the additive gap which is in contrast to BIN PACKING, where such a lower bound is not known.

- We show that SET COVER WITH ORDERED REPLACEMENT does not allow an *asymptotic polynomial time approximation scheme (APTAS)*. Also this distinguishes SET COVER WITH ORDERED REPLACEMENT from BIN PACKING.

- We show that the multiplicative gap of SET COVER WITH ORDERED REPLACEMENT is $\Theta(\log \log n)$. Also this is in sharp contrast to BIN PACKING, where the multiplicative gap is constant.

- Finally we provide a quasi-polynomial (running time $n^{O(\log n)}$) factor 2 approximation algorithm for SET COVER WITH ORDERED REPLACEMENT.

We also bound the additive integrality gap in the case where the replacement order is not a total order. Recall that the *Dilworth number* of a partial order is the smallest number of disjoint chains that cover all elements. Let us denote the additive integrality gap as:

$$\mathrm{gap}(n,d,k) = \max_{\mathcal{S},w} \{OPT(\mathcal{S}) - OPT_f(\mathcal{S})\},$$

where the $\max$ ranges over all set systems over ground set $[n]$ (and proper cost function $w : \mathcal{S} \to [0,1]$), that respect a partial order with Dilworth number $d$ and contain sets of size at most $k$.[2] We show that $\mathrm{gap}(n,d,k) = O(d^2 \log k \log^2 n)$. Our result is an *algorithmic* result in the following sense. If the linear programing relaxation of (1) can be efficiently solved, then a solution of (1) respecting the additive gap can be efficiently computed.

---
[2] We sometimes abbreviate $\mathrm{gap}(n,d) := \mathrm{gap}(n,d,n)$ for systems without cardinality restrictions and $\mathrm{gap}(n) := \mathrm{gap}(n,1)$ if the partial order is total.

We furthermore demonstrate the applicability of our bounds on the additive gap by providing an extensive list of problems from the literature that can be modeled as SET COVER WITH ORDERED REPLACEMENT.

**Related work**

We discuss many BIN PACKING variants that are covered by our framework in the appendix. For many of these problems, there exist *asymptotic polynomial time approximation schemes (APTAS)* [11, 6, 3, 4, 1] or *asymptotic fully polynomial time approximation schemes (AFPTAS)* [16, 8, 7, 9]. An AFPTAS for problem (1) is a polynomial time algorithm (in $n$ and $1/\varepsilon$) that, given an $\varepsilon > 0$ computes a solution $APX(\mathcal{S})$ with
$$APX(\mathcal{S}) \leq (1+\varepsilon)OPT(\mathcal{S}) + f(\varepsilon)$$
where $f$ is a fixed function. This function $f$ can be even exponential in $1/\varepsilon$, see, e.g. [8, 9]. While our additive gap result is incomparable with the quality achieved by an AFPTAS it however sheds some light on how large $n$ has to be in order for such an AFPTAS to be competitive with our result in combination with a simple constant factor approximation algorithm.

We have an additive $O(\log^3 n)$ bound for SET COVER WITH ORDERED REPLACEMENT. If we are considering a particular family of instances with $OPT(\mathcal{S}) \geq \log^4(n)$ for each instance $\mathcal{S}$, then this yields an AFPTAS with $f(\varepsilon) = O((1/\varepsilon)^3)$. Suppose now that $OPT(\mathcal{S}) \leq \log^4(n)$ and suppose that there exists an AFPTAS with an exponential $f$ say $f(\varepsilon) = 2^{1/\varepsilon}$. Then the dimension $n$ has to be *doubly exponential* in $1/\varepsilon$ before the AFPTAS starts to beat the quality of a factor 2 approximation algorithm.

We would also like to mention recent related work on the additive gap of BIN PACKING. The paper [5] relates a prominent conjecture of Beck from the field of *discrepancy theory* to the question whether the additive gap of BIN PACKING is constant. If Beck's conjecture holds true, then the BIN PACKING gap is constant for 3-partitioning instances. While essentially all results on integrality gaps for BIN PACKING variants in the literature use the sparse support of basic solutions, [20] provides bounds based on probabilistic techniques. The work in [20] provides better additive gaps for example in case of BIN PACKING WITH REJECTION.

## 2 Bounding the additive gap

In this section we provide upper and lower bounds for the additive integrality gap of SET COVER WITH ORDERED REPLACEMENT. The upper bound in case of a total order is $O(\log^3 n)$ while the lower bound is of order $\Omega(\log n)$. This result shows that in order to have a polylogarithmic additive integrality gap, it is sufficient to have a total ordering on the elements, like for classical BIN PACKING.

### 2.1 The upper bound

We first deal with an upper bound on the additive integrality gap. We recall that $d$ is the Dilworth number of the partial order, $n$ denotes the number of elements and $k$ is an upper bound on the cardinality of the sets in the family $\mathcal{S}$. We show the following general theorem.

**Theorem 1.** *One has* $gap(n, d, k) = O(d^2 \log k \log^2 n)$.

As in [16], we will construct an integer solution to (1) from a fractional one with the claimed additive bound, by doing a sequence of iterations. At each iteration, we will cover part of our elements by rounding down an optimal fractional solution of a proper updated linear program, modify the residual instance, and re-optimize.

More precisely, we will consider the following (more general) linear program
$$\min\Big\{ \sum_{S \in \mathcal{S}} w(S) x_S \mid \sum_{S \in \mathcal{S}} x_S \cdot \chi(S) \geq b, x \geq \mathbf{0} \Big\}, \tag{2}$$

where $b \in \mathbb{N}_0^n$ is a non-negative vector. The number $b_i$ denotes the *multiplicity* of the item $i$, i.e., how many times it needs to be covered. The reason of having multiplicity is that during our iterations, we will reduce the number of constraints of the above linear program at the expense of increasing multiplicity for a subset of the items. Of course, when multiplicity comes into play, we allow $S \in \mathcal{S}$ to a be multiset, since e.g. if a set $S' = \{i, j, h\} \in \mathcal{S}$ and $i \preceq j \preceq h$, the replacement order implies $S = \{i, i, i\}$ to be in $\mathcal{S}$ as well.

Let us denote the optimum of this linear program with multiplicity and corresponding integer program by $OPT_f(\mathcal{S}, b)$ and $OPT(\mathcal{S}, b)$ respectively. As in [16] we will consider an optimal vertex solution $x^*$ of the linear program (2) and extract the partial covering $\lfloor x_S^* \rfloor$, $S \in \mathcal{S}$. This partial covering will leave some elements of the multiset uncovered. This multiset of uncovered elements defines a *residual instance*, encoded by $b' = \sum_{S \in \mathcal{S}} \{x_S^*\} \cdot \chi(S)$ where $\{x_S^*\} = x_S^* - \lfloor x_S^* \rfloor$ denotes the fractional part of $x^*$. The following relation holds even for arbitrary set covering problems:

$$OPT(\mathcal{S}, b) - OPT_f(\mathcal{S}, b) \leq OPT(\mathcal{S}, b') - OPT_f(\mathcal{S}, b'). \tag{3}$$

The key point of considering a residual instance, is that we will modify it to reduce the number of constraints in the next iteration by using a grouping technique similar to the one given for BIN PACKING in [16]. However, the grouping in [16] crucially relies on the *size of an element* which is part of a BIN PACKING instance, and that is instead missing in our abstract setting. In order to still apply grouping techniques, we will define *pseudo sizes* $s_i \in ]0, 1]$ for each element $i$ in our ground set below. These pseudo sizes satisfy the following properties:

(i) $s_i \geq s_j$ if $i \succeq j$;
(ii) We can cover all the elements of any (not necessarily maximal) chain $C$ at cost $O(\sum_{i \in C} s_i) + O(\log \frac{1}{s_{\min}})$, where[3] $s_{\min} := \min_{i \in [n]} s_i$.

Notice that the usual size of an element in a BIN PACKING instance is also a pseudo size.[4]

Given a vector of pseudo sizes $s$, we can define the *total size* of the elements by $\text{size}(b) = b^T s$. Suppose now that the largest total pseudo size of a set in $\mathcal{S}$ is $\alpha = \max\{\sum_{i \in S} s_i \mid S \in \mathcal{S}\}$. An important observation is that the total size of the residual instance is bounded by

$$\text{size}(b') = b'^T s = \sum_{S \in \mathcal{S}} \{x^*(S)\} \chi(S)^T s \leq \text{support}(b) \cdot \alpha, \tag{4}$$

where $\text{support}(b)$ denotes the number of nonzero entries in $b$.

We are now ready to state the following Lemma, which generalizes the result of Karmakar and Karp [16].

**Lemma 2.** *Let $\mathcal{S}$ be a set system on $[n]$ with cost function $w : \mathcal{S} \to [0, 1]$, $\preceq$ be a partial order respected by $\mathcal{S}$ of Dilworth number $d$ and let $b \in \mathbb{N}_0^n$. Still, let $s$ be a vector of pseudo sizes which satisfies properties (i) and (ii). Then $OPT(\mathcal{S}, b) - OPT_f(\mathcal{S}, b) = O(\alpha \log(1/s_{\min}) \cdot d \log n)$, where $\alpha = \max\{\sum_{i \in S} s_i \mid S \in \mathcal{S}\}$.*

*Proof.* Initially, we partition the elements into chains $C_1, \ldots, C_d$ w.r.t. the order $\succeq$. The bound follows from iterating the following procedure. First, we replace $b$ by $b'$ which encodes its residual instance. According to (3), this does not decrease the additive integrality gap. We then apply the following grouping procedure to each chain $C_\mu$ with $\mu = 1, \ldots, d$. The chain $C_\mu$ is partitioned into *classes*:

$$U_\ell^\mu = \left\{ i \in C_\mu \mid \left(\frac{1}{2}\right)^{\ell+1} < s_i \leq \left(\frac{1}{2}\right)^\ell \right\} \text{ for } \ell = 0, \ldots, \lfloor \log(1/s_{\min}) \rfloor.$$

For each such class $U_\ell^\mu$, build groups of $4 \cdot 2^\ell \alpha$ consecutive elements (the elements are always counted with multiplicity), starting from the largest element (the last group could contain less elements), and discard the first and the last group. In this way, we discard at most $8 \cdot 2^\ell \cdot \alpha$ elements in the class $U_\ell^\mu$. Those elements have total size at most $8 \cdot \alpha$, hence the total size of discarded elements in chain $C_\mu$ is bounded by $8 \cdot \alpha \cdot (\log(1/s_{\min})+1)$. By (ii) we can cover them at cost $O(\alpha \cdot \log(1/s_{\min}))$. This amounts to a cost of $O(d \cdot \alpha \cdot \log(1/s_{\min}))$ to cover all discarded elements of all chains.

---
[3]For notational convinience, we always assume that $s_{\min} \leq 1/2$ so that $\log \frac{1}{s_{\min}} \geq 1$.
[4]Here one can cover a subset of elements even with $2 \sum_{i \in C} s_i + 1$ bins alone and the minimal size does not need to be considered.

Then, we "round-up" the elements in each group to the largest element in this group. In other words, for each group we now consider to have one item type (the largest one, according to the chain) with multiplicity $4 \cdot 2^\ell \alpha$. This way, we obtain a new "rounded" instance that is represented by a vector $b'' \in \mathbb{N}_0^n$. Since the discarded groups compensate the round-up operation within each group, one has $OPT_f(\mathcal{S}, b'') \leq OPT_f(\mathcal{S}, b')$. Also, $OPT(\mathcal{S}, b') \leq OPT(\mathcal{S}, b'') + O(d \cdot \alpha \cdot \log(1/s_{\min}))$ and thus

$$OPT(\mathcal{S}, b') - OPT_f(\mathcal{S}, b') \leq OPT(\mathcal{S}, b'') - OPT_f(\mathcal{S}, b'') + O(d \cdot \alpha \cdot \log(1/s_{\min})).$$

We will next show that $\text{support}(b'') \leq \text{support}(b)/2$. The assertion then follows, since the support of the rounded instance (and hence the corresponding additive integrality gap) will be 1 after $O(\log n)$ iterations of the above described procedure.

The support of $b''$ is the number of non-discarded groups. Each non-discarded group $U_\ell^\mu$ contains a number of elements equal to $4 \cdot 2^\ell \alpha$, each of size at least $\frac{1}{2^{\ell+1}}$. Then the total size of the group is at least $2 \cdot \alpha$. Thus $2 \cdot \alpha \cdot \text{support}(b'') \leq \text{size}(b')$. But since $b'$ encodes a residual instance, one has $\text{size}(b') \leq \text{support}(b) \cdot \alpha$ from (4). That is, $\text{support}(b'') \leq \text{support}(b)/2$ follows. □

Notice that the $O(\log^2 n)$ upper bound of Karmarkar and Karp [16] for BIN PACKING also follows from this lemma by considering the original sizes given in the instances as pseudo sizes, and setting $d$, $\alpha$ and all the initial $b_i$'s to one. Items of size less than $1/n$ can be removed and distributed on top of the solution later. If one needs an additional bin, the gap is even bounded by a constant.

We now show how to define good pseudo sizes of the items for *any* given set system $\mathcal{S}$. Let

$$\text{size}(i) := \min\left\{ \frac{w(S)}{|S|} \mid S \in \mathcal{S} \text{ contains only elements } j \succeq i \right\}$$

Note that $\text{size}(i) \geq \text{size}(j)$ holds if $i \succeq j$, that is, property (i) of Lemma 2 holds. The next Lemma shows that also (ii) holds as well.

**Lemma 3.** *Let $\mathcal{S}$ be a set system with replacement property. We can cover all the elements of a (not necessarily maximal) chain $C$ at cost at most $2 \sum_{i \in C} \text{size}(i) + O(\log \frac{1}{s_{\min}})$.*

*Proof.* Again let $U_\ell = \{i \in C \mid (\frac{1}{2})^{\ell+1} < \text{size}(i) \leq (\frac{1}{2})^\ell\}$ be the $\ell$th size class of chain $C$. We construct a solution for each size class separately. For class $\ell$, consider iteratively the largest uncovered element $i$ and let $S(i)$ be a set, defining the quantity $\text{size}(i)$, i.e. $S(i)$ contains only elements that are at least as large as $i$ and $\frac{w(S(i))}{|S(i)|} \leq \text{size}(i)$. By the replacement property, this set $S(i)$ can be used to cover the largest $|S(i)|$ many uncovered elements.

Let $S_1, \ldots, S_p$ be the sets selected in this procedure. Note that all the $S_j$, but possibly $S_p$, cover exactly $|S_j|$ elements. Then, since $\frac{w(S_j)}{|S_j|} \leq 2\text{size}(i)$ for every $i \in S_j$, we have

$$\sum_{j=1}^p w(S_j) = \sum_{j=1}^{p-1} \sum_{i \in S_j} \frac{w(S_j)}{|S_j|} + w(S_p) \leq \sum_{i \in U_\ell} 2 \cdot \text{size}(i) + 1.$$

Summing over the $(\log \frac{1}{s_{\min}} + 1)$ many size classes then gives the claim. □

We are ready to prove Theorem 1.

*Proof of Theorem 1.* Let $\mathcal{S}$ be a set system on $[n]$, and $\preceq$ be a partial order respected by $\mathcal{S}$. We may assume that $\preceq$ consists of $d$ incomparable chains $C_1, \ldots, C_d$. We define $s_i = \text{size}(i)$ for any element $i$.

We first claim that we can discard all items $i$ with $s_i < \frac{1}{n^2}$: in fact, by the definition of the pseudo sizes any such element is contained in a set of cost at most $\frac{1}{n}$, hence all such tiny elements can be covered at a total cost of 1.

Let $S \in \mathcal{S}$ and consider an element $i$ which is the $j$th largest in $S \cap C_\mu$. By the definition of the pseudo sizes, $\text{size}(i) \leq \frac{w(S)}{j}$. Thus

$$\sum_{i \in S} \text{size}(i) \leq d \sum_{j=1}^{k} \frac{w(S)}{j} \leq 2d \log k.$$

Therefore, it follows from Lemma 2 (by setting $b = \mathbf{1}$) that $\text{gap}(n, d, k) = O(d^2 \log k \log^2 n)$, since $\alpha = 2d \log k$ and $s_{\min} \geq 1/n^2$. $\square$

We want to remark that the above result is constructive in the sense that once a fractional solution $x$ and the partial order are given, a feasible integer solution matching the bound of Theorem 1 can be computed in polynomial time.

If all costs are one, i.e., $w(S) = 1$ for each $S \in \mathcal{S}$, then we have $\text{size}(i) \geq \frac{1}{k}$. Thus we have the following corollary.

**Corollary 4.** *If all costs of feasible sets are one, one has $\text{gap}(n, d, k) = O(d^2 \log^2 k \log n)$.*

When $d = 1$, this theorem says $\text{gap}(n, 1, k) = O(\log^2 k \log n)$, which is better than the result of [5]: $\text{gap}(n, 1, k) = O(k \log n)$.

### 2.2 The lower bound

In this section, we give a lower bound for the additive integrality gap of set systems with replacement property. For simplicity, we first assume that $\preceq$ consists of only one chain.

**Lemma 5.** *One has $\text{gap}(n) \geq \Omega(\log n)$.*

*Proof.* Let $m$ be a parameter, that we determine later. Define a unit cost set system $\mathcal{S}$ as follows. For any $\ell \in \{1, \ldots, m\}$, introduce $3 \cdot 100^\ell$ many $\ell$-*level elements* $U_\ell$. Hence $U := \bigcup_{\ell=1}^{m} U_\ell$ with $|U_\ell| = 3 \cdot 100^\ell$ is the set of all elements. We choose $U_1 \succeq U_2 \succeq \ldots \succeq U_m{}^5$ and an arbitrary order within every $U_\ell$. Any set $S$ of at most $2 \cdot 100^\ell$ many $\ell$-level elements forms an $\ell$-level set, e.g.

$$\mathcal{S} = \bigcup_{\ell=1}^{m} \left\{ S \subseteq U_\ell \cup \ldots \cup U_m \mid |S| \leq 2 \cdot 100^\ell \right\}.$$

By taking 3 $\ell$-level sets to the fractional extend of $\frac{1}{2}$, we can cover all $\ell$-level elements, hence $OPT_f(\mathcal{S}) \leq \frac{3}{2} \cdot m$. It remains to lower bound $OPT(\mathcal{S})$. Let $n_\ell$ be the number of $\ell$-level sets in any integer solution. To cover all level $\ell$-level elements we must either have $n_\ell \geq 2$, or

$$\sum_{j < \ell} n_j \cdot 2 \cdot 100^j \geq 100^\ell \Leftrightarrow 4 \sum_{j < \ell} n_j \cdot 100^{j-\ell} \geq 2$$

In any case, the sum of the left hand sides must be at least 2, i.e.

$$n_\ell + 4 \sum_{j < \ell} n_j \cdot 100^{j-\ell} \geq 2 \tag{5}$$

We add up (5) for $\ell = 1, \ldots, m$ and obtain $(*)$

$$\sum_{\ell=1}^{m} n_\ell \cdot \underbrace{\left(1 + 4 \sum_{i \geq 1} \frac{1}{100^i}\right)}_{= \frac{103}{99}} \geq \sum_{\ell=1}^{m} \left(n_\ell + 4 \sum_{j < \ell} n_j \cdot 100^{j-\ell}\right) \overset{(*)}{\geq} 2m$$

---
[5]We abbreviate $U \succeq U' \Leftrightarrow \forall i \in U, j \in U' : i \succeq j$.

This gives

$$OPT(\mathcal{S}) - OPT_f(\mathcal{S}) \geq \sum_{\ell=1}^{m} n_\ell - \frac{3}{2}m \geq \underbrace{2\frac{99}{103}m}_{>1.9m} - \frac{3}{2}m > 0.4 \cdot m$$

The number of elements in the instance is $n = \sum_{\ell=1}^{m} 3 \cdot 100^\ell$, hence $m = \Omega(\log n)$ and the claim follows. □

More general, the following holds:

**Theorem 6.** $gap(n, d) \geq \Omega(d \cdot \log(n/d))$.

*Proof.* Apply the construction from Lemma 5 to obtain families $\mathcal{S}_1, \ldots, \mathcal{S}_d$, each on a disjoint set of $n/d$ elements. Then the union $\bigcup_{j=1}^{d} \mathcal{S}_j$ has Dilworth number $d$ and the claimed additive gap. □

## 3 Approximation & intractability

In the appendix, we mention many variants of BIN PACKING that admit an APTAS (or even an AFPTAS) and fall into our general framework of SET COVER WITH ORDERED REPLACEMENT. It is thus natural to ask whether the SET COVER WITH ORDERED REPLACEMENT itself has an APTAS. This cannot be expected for arbitrary partial orders. In this section we show that an APTAS does not exist, even if the replacement order is a total order. On the positive side however, we show that there exists a quasi polynomial time factor 2 approximation algorithm in this case.

From now on, we restrict our view exclusively on set systems, respecting a total order $\preceq$. To define such a set system with unit-costs, it will be convenient to consider just the set of *generators*. These are the sets that are maximal with respect to the order. More formally, if there is an injective map $\varphi : S \to S'$ with $i \preceq \varphi(i)$ for all $i \in S$ (i.e. we can obtain a set $S$ from $S'$ by applying the replacement rule), then we say that $S'$ *dominates* $S$. Hence a set family $\mathcal{S}$ is called the set of *generators* for the set system

$$g(\mathcal{S}) = \{S \subseteq [n] \mid \exists S' \in \mathcal{S} : S' \text{ dominates } S\}$$

Hence, if $\mathcal{S}$ is an arbitrary set family, by definition $g(\mathcal{S})$ respects the replacement rule. It is not difficult to see that the following holds (see the Appendix for a formal proof).

**Proposition 7.** Let $\mathcal{S} \subseteq 2^{[n]}$ be a family of sets and $\succeq$ the total order with $1 \succeq 2 \succeq \ldots \succeq n$.

  i) If $\mathcal{S}' \subseteq g(\mathcal{S})$ is a feasible solution (i.e. $\bigcup_{S \in \mathcal{S}'} S = [n]$) then $\sum_{S \in \mathcal{S}'} |S \cap \{1, \ldots, i\}| \geq i$ for $i = 1, \ldots, n$.

  ii) If $\mathcal{S}' \subseteq \mathcal{S}$ are generators with $\sum_{S \in \mathcal{S}'} |S \cap \{1, \ldots, i\}| \geq i \ \forall i \in [n]$, then sets $\mathcal{S}'$ can be replaced by dominated ones which form a feasible solution of the same cardinality.

### 3.1 Ruling out an APTAS

We will now see that unless $\mathbf{P} = \mathbf{NP}$, there is no APTAS for a generic problem defined on a set system that respects a total order.

**Theorem 8.** *For every $\varepsilon > 0$ and any $C > 0$ there is a generic problem with unit-cost sets respecting a total order for which it is $\mathbf{NP}$-hard to find a solution of cost $(\frac{3}{2} - \varepsilon)OPT + C$.*

*Proof.* We will prove the theorem by constructing a set system such that for any fixed integer $k > 0$, it is $\mathbf{NP}$-hard to distinguish whether an optimum solution consists of at most $2k$ or at least $3k$ sets. Choosing $k := k(\varepsilon, C)$ large enough then gives the claim.

To establish this hardness result, we will use a reduction from the $\mathbf{NP}$-hard PARTITION [12] problem. An instance $\mathcal{I}$ of PARTITION, is given by a set of $n$ items with sizes $a_1 \geq a_2 \geq \ldots \geq a_n$, and asks for a partition of the items into two sets $A_1$ and $A_2$, such that $\sum_{j \in A_1} a_j = \sum_{j \in A_2} a_j =: A$. Given such an instance, we create groups $L_1, \ldots, L_k$, where the group $L_p$ contains $n^{2p}$ copies of item $i$, i.e. $L_p = \{v_{p,i}^j \mid i = 1, \ldots, n; j = 1, \ldots, n^{2p}\}$. Note that the total number of elements is $N := n^{O(k)}$. We define a total order with $L_1 \succeq \ldots \succeq L_k$ and $v_{p,i}^j \succeq v_{p,i'}^{j'}$ whenever $i < i'$ (and $v_{p,i}^1 \succeq \ldots \succeq v_{p,i}^{n^{2p}}$ for the sake of completeness). Let $S(I, p) := \{v_{p,i}^j \mid i \in I; j = 1, \ldots, n^{2p}\}$ be the set induced by $I \subseteq [n]$ in group $L_p$. We define generators

$$\mathcal{S} = \left\{ S(I, p) \mid \forall p = 1, \ldots, k;\ \forall I \subseteq [n] : \sum_{i \in I} a_i \leq A \right\}$$

*Completeness:* $\mathcal{I} \in \text{PARTITION} \Rightarrow OPT(g(\mathcal{S})) \leq 2k$. Let $I \subseteq [n]$ with $\sum_{i \in I} a_i = A$ be a PARTITION solution. Then the $2k$ sets of the form $S([n] \setminus I, p), S(I, p)$ for $p = 1, \ldots, k$ cover all $N$ elements.

*Soundness:* $\mathcal{I} \notin \text{PARTITION} \Rightarrow OPT(g(\mathcal{S})) \geq 3k$. We may assume that $3k < n$. Now suppose for the sake of contradiction, there is no PARTITION solution, but $\mathcal{S}' \subseteq \mathcal{S}$ is a family of less than $3k$ generating sets, dominating a feasible solution, satisfying the second condition in Prop. 7. Then there must be a group $L_p$ such that $\mathcal{S}'$ contains generators $S(I_1, p), \ldots, S(I_m, p)$ with $m \leq 2$. Then from Prop. 7 we obtain that

$$i \cdot n^{2p} \stackrel{\text{Prop. 7}}{\leq} \sum_{S \in \mathcal{S}'} |S \cap (L_1 \cup \ldots \cup L_{p-1} \cup S(\{1, \ldots, i\}, p))| \leq 3k \cdot n \cdot n^{2p-2} + \sum_{\ell=1}^m n^{2p} |I_\ell \cap \{1, \ldots, i\}|$$

Hence

$$\sum_{\ell=1}^m |I_\ell \cap \{1, \ldots, i\}| \geq \left\lceil i - \frac{3k}{n} \right\rceil = i$$

since the left hand side is integral and $3k < n$. Since $\sum_{\ell=1}^m |I_\ell \cap \{1, \ldots, i\}| \geq i$ for all $i \in [n]$, we conclude by applying again Prop. 7, that elements in $I_1, \ldots, I_m$ can be replaced by smaller ones and obtain $I_1', \ldots, I_m'$ such that still $\sum_{i \in I_\ell'} a_i \leq A$ but $\bigcup_{\ell=1}^m I_\ell' = [n]$. Since $m \leq 2$, this is a contradiction. □

### 3.2 A 2-approximation in quasi-polynomial time

A FEASIBILITY ORACLE for $\mathcal{S}$ is a polynomial time algorithm, which decides for an input set $S \subseteq [n]$, whether $S \in \mathcal{S}$ and if the answer is yes, it reports the cost $w(S)$.

**Theorem 9.** *Let $\mathcal{S} \subseteq 2^{[n]}$ be a set system, respecting a total order, having a cost function $w : \mathcal{S} \to [0, 1]$ and admitting a FEASIBILITY ORACLE. Then one can compute a feasible solution $\mathcal{S}' \subseteq \mathcal{S}$ of cost $\sum_{S \subseteq \mathcal{S}'} w(S) \leq 2 \cdot OPT(\mathcal{S})$ in time $n^{O(\log n)}$.*

*Proof.* We partition the elements into groups $L_0, \ldots, L_k$ of elements such that $L_0 \succeq \ldots \succeq L_k$, $|L_i| = 2^i$ for all $i < k$ (group $L_k$ might have fewer elements). Note that $k \leq \log n$. Round up all items in $L_i$ to the largest one in $L_i$ and term the new instance $I'$. In other words, in the instance $I'$ we have one item type for each group $i$ (the largest one according to the total order) with multiplicity $2^i$. Buying the sets in $OPT(\mathcal{S})$ twice is sufficient to cover all elements in the rounded instance $I'$. The new instance has just $k + 1 \leq \log n + 1$ many different item types. Hence, let $\mathcal{P} = \{p \in \mathbb{Z}_+^{k+1} \mid \text{set } S(p) \text{ where } i \text{ has multiplicity } p_i \text{ is in } \mathcal{S}\}$ be the set of feasible patterns, and define the cost of $p$ to be equal to the cost of the corresponding set $S(p)$, i.e. $w(p) = w(S(p))$. Still let $b$ be the vector of item multiplicities. Then $|\mathcal{P}| \leq (n+1)^{(\log n + 1)}$. Next, consider table entries

$$A(b') = \min\left\{ \sum_{p \in \mathcal{P}} w(p) x_p \mid \sum_{p \in \mathcal{P}} x_p \cdot p = b',\ x \in \mathbb{Z}_+^\mathcal{P} \right\}$$

Then such table entries can be computed by dynamic programming in time $n^{O(\log n)}$, as follows:

- Initialize $A(b') = 0$ for all $b' \leq \mathbf{0}$;
- Compute $A(b') = \min_{p \in \mathcal{P}} \{A(b' - p) + w(p)\}$.

The entry $A(b)$ yields $OPT(I')$. The corresponding solution can be easily reconstructed as well. □

# 4 Multiplicative integrality gaps

In this section, we show that in case of a total order, the multiplicative integrality gap is $\Theta(\log \log n)$.

**Lemma 10.** *Let $\mathcal{S}$ any set family on $n$ elements respecting a total order and let $w : \mathcal{S} \to [0,1]$ be a cost function. Then $OPT(\mathcal{S}) \leq O(\log \log n) \cdot OPT_f(\mathcal{S})$.*

*Proof.* Consider consecutive groups $L_0, \ldots, L_k$ such that $|L_i| = 2^i$ (group $L_k$ might have fewer elements), $k \leq \log n$ and all elements in $L_i$ are larger than those in $L_{i+1}$. Let $x \in [0,1]^\mathcal{S}$ be a fractional solution. We buy set $S$ independently with probability $\lambda \cdot x_S$ where $\lambda := \max\{8 \cdot \log(4 + 4 \log n), 4\}$ (in fact we may assume that $\lambda \cdot x_S \leq 1$, otherwise we buy $\lfloor \lambda x_S \rfloor$ sets deterministically and then another set with probability $\lambda x_S - \lfloor \lambda x_S \rfloor$). Let $X_S$ be the indicator variable, telling whether we bought $S$ or not. Define $E_i := \{\sum_S X_S \cdot |S \cap L_i| < 2 \cdot |L_i|\}$ as the event that we bought less than two times enough sets for the $i$th group. Recall the following Chernov bound (see[6] e.g. [18]):

*Let $Y_1, \ldots, Y_m$ be independent random variables with $Y_i \in [0,1]$, $Y = \sum_{i=1}^m Y_i$ and $0 < \delta < 1$. Then $\Pr[Y \leq (1-\delta)E[Y]] \leq e^{-E[Y]\delta^2/2}$.*

Applying the Chernov bound with $\delta := 1/2$, $Y_S := X_S \frac{|S \cap L_i|}{|L_i|}$ and $E[Y] = \lambda$, we obtain

$$\Pr[E_i] = \Pr\left[\sum_{S \in \mathcal{S}} X_S \cdot |S \cap L_i| < 2 \cdot |L_i|\right] \leq \Pr\left[\sum_{S \in \mathcal{S}} Y_S < (1-\delta)\lambda\right] \leq e^{-\lambda/8} \leq \frac{1}{4(1 + \log n)}.$$

By the union bound, $\Pr[E_0 \cup \ldots \cup E_k] \leq \frac{1}{4}$. Furthermore $\Pr[\sum_{S \in \mathcal{S}} X_S w(S) > 4\lambda \cdot OPT_f] \leq \frac{1}{4}$ by Markov's inequality. Overall, with probability at least $1/2$, we obtain an integer solution $\mathcal{S}' = \{S \in \mathcal{S} \mid X_S = 1\}$ of cost at most $O(\log \log n) \cdot OPT_f$ that reserves at least $2|L_i|$ slots for elements in $L_i$. Those slots are enough to cover all elements in $L_{i+1}$. □

Note that the result in Lemma 10 provides a randomized polynomial time algorithm, provided that a near-optimal fractional solution $x$ can be obtained. We now state a tight lower bound on the multiplicative integrality gap.

**Lemma 11.** *There exists a set system $\mathcal{S}' \subseteq 2^{[n]}$ with unit cost sets and respecting a total order such that $OPT(\mathcal{S}') \geq \Omega(\log \log n) \cdot OPT_f(\mathcal{S}')$.*

*Proof.* Let $k \in \mathbb{N}$ be a parameter. To construct our instance, we use as starting point a SET COVER instance defined by a set system $\mathcal{C}$ with $2^k - 1$ sets $C_1, \ldots, C_{2^k-1}$ and $2^k - 1$ elements $U = \{1, \ldots, 2^k - 1\}$ such that one needs at least $k$ sets to cover all elements, while $OPT_f(\mathcal{C}) \leq 2$ (see Example 13.4 in the book of Vazirani [21]).

For every element $i \in U$, create *groups* $L_i$ with $|L_i| = (2k)^i$. For any $C_j$ in the original set system, define a set $S_j := (\bigcup_{i \in C_j} L_i)$ with unit cost, and let $\mathcal{S} = \{S_1, \ldots, S_{2^k-1}\}$. In other words, we take a standard SET COVER instance and replace the $i$th element by $(2k)^i$ elements. We define a total order such that all items in $L_i$ are larger than those in $L_{i+1}$ (and any order within the groups). The claim is that the set system $\mathcal{S}' := g(\mathcal{S})$, which is generated by the sets $\mathcal{S}$, has a covering integrality gap of at least $k/2$. First note that still $OPT_f(\mathcal{S}') \leq 2$. Now suppose for contradiction that there are generators (after reindexing) $S_1, \ldots, S_m \subseteq \mathcal{S}$ with $m < k$ that satisfy the condition in Prop. 7. Since $m < k$, there must be an index $i$ such that $i \notin (C_1 \cup \ldots \cup C_m)$. Then

$$(2k)^i \leq \sum_{\ell=1}^i |L_\ell| \stackrel{\text{Prop. 7}}{\leq} \sum_{j=1}^m |S_j \cap (L_1 \cup \ldots \cup L_i)| = \sum_{j=1}^m \sum_{\ell \in C_j, \ell \leq i} (2k)^\ell \stackrel{i \notin C_j}{\leq} m \cdot 2(2k)^{i-1} = \frac{m}{k}(2k)^i$$

Rearranging yields that $m \geq k$. Note that the number of elements in the system $\mathcal{S}'$ is $n = \sum_{i=1}^{2^k-1}(2k)^i \leq 2 \cdot (2k)^{2^k}$, hence $k = \Omega(\log \log n)$. □

---

[6]To be precise, the claim in [18] is for $0/1$ distributed random variables, but the same proof goes through if $0 \leq Y_i \leq 1$.

# A  Applications

We now demonstrate the versatility of our framework by establishing small additive integrality gaps for several BIN PACKING variants from the literature.

In the following, whenever we apply Lemma 2 we implicitly consider $b = \mathbf{1}$.

## Cardinality Bin Packing

For CARDINALITY BIN PACKING one is given a BINPACKING instance $s_1, \ldots, s_n$ with an additional parameter $k \in \mathbb{N}$. A subset of items $S$ can be assigned to a bin only if $\sum_{i \in S} s_i \leq 1$ and $|S| \leq k$. There is an AFPTAS due to Epstein & Levin [8].

This problem is a special case of a set system with total replacement order, and hence $\mathrm{gap}(n, 1, k)$ upper bounds the additive integrality gap. Using Lemma 2 we can have a better gap.

**Corollary 12.** *Let $\mathcal{S}$ be a* CARDINALITY BIN PACKING *instance. Then* $OPT(\mathcal{S}) - OPT_f(\mathcal{S}) = O(\log k \log n)$.

*Proof.* Define pseudo sizes $a_i$ to be $a_i = \max\{s_i, 1/k\}$ for each item $i$. The $a_i$'s clearly respect property (i) of Lemma 2. We now show that also property (ii) is satisfied.

Let $C$ be any subset of the items. Apply the First-Fit algorithm for BINPACKING (see [14]) with respect to $a_i$'s, which works as follows: Consider an uncovered item $i$. Take any open bin $B : \sum_{j \in B} a_j \leq 1 - a_i$, and assign $i$ to $B$. If no such bin exists, open a new bin and assign $i$ to it. Repeat the process till all the elements are covered. Note that this algorithm returns sets $S$ which are feasible for the original BIN PACKING instance. In particular, any open bin (but possibly the last one) satisfies: $\sum_{i \in B} a_i \geq \frac{1}{2}$. That is, we can cover the items at cost at most $2 \sum_{i \in C} a_i + 1$.

Still, since $\sum_{i \in S} a_i \leq 2$ for any $S \in \mathcal{S}$, Lemma 2 implies that $OPT(\mathcal{S}) - OPT_f(\mathcal{S}) = O(\log k \log n)$. □

## Open End Bin Packing

In OPEN END BIN PACKING a standard BIN PACKING instance $s_1, \ldots, s_n$ is given, but a set $S \subseteq [n]$ is feasible if $\sum_{i \in S \setminus \{j\}} s_i \leq 1$ holds for every $j \in S$, i.e. discarding any item (in particular the smallest one) from the bin, brings the size down to at most 1. There is an FPTAS for this variant by Epstein & Levin [7]. Still in this notion, note that replacing an item by a smaller one, leaves a set feasible, i.e. the feasible sets have the replacement property. Using Lemma 2 we can obtain the same additive integrality gap as BINPACKING problem.

**Corollary 13.** *Let $\mathcal{S}$ be a* OPEN END BIN PACKING *instance. Then* $OPT(\mathcal{S}) - OPT_f(\mathcal{S}) = O(\log^2 n)$.

*Proof.* Since a solution of the BINPACKING instance with item sizes $s_1, \ldots, s_n$ is also feasible for OPEN END BIN PACKING, the First-Fit algorithm for BINPACKING implies that we can cover any subset $C$ of the items at cost at most $2 \sum_{i \in C} s_i + 1$. Moreover, it holds that $\sum_{i \in S} s_i \leq 2$ for any $S \in \mathcal{S}$, and we may assume $s_i \geq 1/n$. Therefore, by Lemma 2, we have $OPT(\mathcal{S}) - OPT_f(\mathcal{S}) = O(\log^2 n)$. □

## Bin Packing with General Cost Structure

In BIN PACKING WITH GENERAL COST STRUCTURE we are given a set of $n$ items with sizes $s_1, \ldots, s_n$, and a cost function $f : \{0, \ldots, n\} \to [0, 1]$, which is a monotonically nondecreasing concave function with $f(0) = 0$. The goal is to find sets $S_1, \ldots, S_p$ to cover the items such that for every $j = 1, \ldots, p$ $\sum_{i \in S_j} s_i \leq 1$ and the total cost $\sum_{i=1}^{p} f(|S_i|)$ is minimized. This problem admits an AFPTAS [9].

Again, using Lemma 2 we can prove the same additive integrality gap as BINPACKING.

**Corollary 14.** *Let $\mathcal{S}$ be a* BIN PACKING WITH GENERAL COST STRUCTURE *instance. Then* $OPT(\mathcal{S}) - OPT_f(\mathcal{S}) = O(\log^2 n)$.

*Proof.* Let $C$ be any subset of the items. Apply the First-Fit algorithm on the items in $C$ with respect to $s_i$'s, and let $\mathcal{S}' \subseteq \mathcal{S}$ be the family of subsets used to cover $C$. We have that $2\sum_{i \in C} s_i + 1 \geq |\mathcal{S}'|$. Since the costs are at most 1, the latter is also an upperbound on the cost payed to cover the items in $C$. Still, $\sum_{i \in S} s_i \leq 1$ for any $S \in \mathcal{S}$, and we may assume $s_i \geq 1/n$. Therefore, by Lemma 2, we can state $OPT(\mathcal{S}) - OPT_f(\mathcal{S}) = O(\log^2 n)$. □

## Generalized Cost Variable Sized Bin Packing

In GENERALIZED COST VARIABLE SIZED BIN PACKING we are given a set of $n$ items with sizes $s_1, \ldots, s_n \in [0, 1]$, and a set of bin types $B_1, \ldots, B_k$ each one with a different capacity $a_1, \ldots, a_k$ and a different cost $c_1, \ldots, c_k \in [0, 1]$. The goal is to select a minimum cost subset of bins to pack the items, where a subset of items can be packed into a bin $B_i$ if the sum of their sizes does not exceed the bin capacity $a_i$. See [6] for an APTAS.

**Corollary 15.** *Let $\mathcal{S}$ be a GENERALIZED COST VARIABLE SIZED BIN PACKING instance. Then $OPT(\mathcal{S}) - OPT_f(\mathcal{S}) = O(\log^3 n)$.*

## Bin Packing with Rejection

For BIN PACKING WITH REJECTION, additionally to the item sizes $s_1, \ldots, s_n$, any item $i$ has a rejection cost $c_i$ (w.l.o.g. $0 < c_i < 1$). An item can either be packed into a bin or discarded at cost $c_i$. In other words, if $\mathcal{S}$ denotes the set of feasible patterns, then

$$OPT_{\text{REJECTION}} = \min_{x \in \{0,1\}^{\mathcal{S}}} \left\{ \mathbf{1}^T x + \sum_{i : i \in S \Rightarrow x_S = 0} c_i \right\}$$

See [8] for an AFPTAS.

To make this problem fit into our framework, let $d := |\{c_i \mid i = 1, \ldots, n\}|$ denote the number of different rejection costs. This can be seen by modelling items with the same cost as a single chain. Feasible patterns become a set of cost 1. Furthermore any $\{i\}$ is a feasible set of cost $c_i$. Then we obtain the following additive integrality gap.

**Corollary 16.** *Let $\mathcal{S}$ be a BIN PACKING WITH REJECTION instance. Then $OPT(\mathcal{S}) - OPT_f(\mathcal{S}) = O(d \log^2 n)$ holds, where $d$ is the number of different costs.*

*Proof.* Since a solution of the BINPACKING instance with item sizes $s_1, \ldots, s_n$ is also feasible for BIN PACKING WITH REJECTION, the First-Fit algorithm for BINPACKING implies that we can cover any subset $C$ of the items at cost at most $2\sum_{i \in C} s_i + 1$. Note that we may assume $s_i \geq 1/n$, and that $\sum_{i \in S} s_i \leq 1$ for any $S \in \mathcal{S}$. Therefore, Lemma 2 implies that $OPT(\mathcal{S}) - OPT_f(\mathcal{S}) = O(d \log^2 n)$. □

Unfortunately, in general, the number of different rejection costs $d$ can be equal to $n$, therefore in this case the bound becomes trivial. On the other hand, we may round the costs such that $c_i \geq \frac{1}{n}$ and $c_i = (1 + \varepsilon)^{\mathbb{Z}}$. Then we have that $d = O(\frac{\log n}{\log(1 + \varepsilon)}) \approx O(\frac{\log n}{\varepsilon})$ for small $\varepsilon$. Therefore

$$OPT(\mathcal{S}) \leq (1 + \varepsilon)(OPT_f(\mathcal{S})) + O(\frac{1}{\varepsilon} \log^3 n).$$

By choosing $\varepsilon = \frac{(\log n)^{3/2}}{\sqrt{n}}$, and considering that $OPT_f(\mathcal{S}) \leq n$:

**Corollary 17.** *Let $\mathcal{S}$ be a BIN PACKING WITH REJECTION instance. Then $OPT(\mathcal{S}) - OPT_f(\mathcal{S}) = O(\sqrt{n} \log^{3/2} n)$.*

We mention that the author in [20] provides a better additive integrality gap for this problem, using probabilistic techniques.

**Train Delivery**

For the TRAIN DELIVERY problem, the input consists of items $i = 1, \ldots, n$, each having a size $s_i$ and position $p_i \in [0, 1]$. We can imagine to have a single train of capacity $1$ with which we have to transport the items to the sink, located at $0$. The goal is to minimize the time, needed for this, if the train has unit speed. In other words, a single train route is a set $S \subseteq [n]$ with $\sum_{i \in S} s_i \leq 1$ and its cost is $w(S) := \max_{i \in S}\{p_i\}$ (omitting a factor of 2). The authors in [4] give an APTAS.

As in the previous paragraph, we may modify the input instance so that $s_i, p_i \geq \frac{1}{n}$. Furthermore we suffer only a $(1+\varepsilon)$-error, if we round the positions, so that $p_i \in (1+\varepsilon)^{\mathbb{Z}}$, hence we have just $O(\frac{\log n}{\varepsilon})$ different positions if $\varepsilon$ is very small. Then we could model items with the same position as a single chain. By choosing $\varepsilon = \frac{(\log n)^{3/2}}{\sqrt{n}}$:

**Corollary 18.** *Let $\mathcal{S}$ be a* TRAIN DELIVERY *instance. Then* $OPT(\mathcal{S}) - OPT_f(\mathcal{S}) = O(\sqrt{n} \log^{3/2} n)$.

**$m$-dimensional Vector Packing**

Let $V$ be a set of $m$-dimensional vectors $v_1, \ldots, v_n \in [0, 1]^m$. The goal is to partition these vectors into bins $B_1, \ldots, B_k$, such that $k$ is minimized and $\sum_{i \in B_j} v_i \leq \mathbf{1}$. Hence for 1-dimensional vectors, this problem is precisely BIN PACKING.

We say that $v_i \preceq v_j$, if $v_j$ is componentwise not smaller than $v_i$. The Dilworth number of $V$ is then the smallest $d$ such that $V$ can be partitioned into subsets $V^1, \ldots, V^d$, having the following property: for every two vectors $v_i$ and $v_j$ in $V^h$, either $v_i \preceq v_j$ or $v_j \preceq v_i$.

Unfortunately, if there is no bound on $d$, there is no APTAS possible already in 2-dimensions [22]. Differently, for constant $d$ there is an APTAS given by Caprara, Kellerer and Pferschy [1]. Since this problem fits into our framework, we have:

**Corollary 19.** *Let $\mathcal{S}$ be a $m$-*DIMENSIONAL VECTOR PACKING *instance, with Dilworth number $d$. Then* $OPT(\mathcal{S}) - OPT_f(\mathcal{S}) = O(d^2 \log^3 n)$.

# B  Proof of Proposition 7

*Proof of proposition 7.*   i) Since $\mathcal{S}'$ covers every item at least once, the first $i$ items are covered at least $i$ many times.

ii) Let $a(j) = \{S \in \mathcal{S}' \mid j \in S\}$ be the number of *slots* that the generators reserve for element $j \in [n]$. Recall that $a_j \in \mathbb{N}_0$ and $\sum_{i \leq j} a(i) \geq j$ by assumption. Define a bipartite graph $G = ([n] \times [n], E)$ with edges $(i, j) \in E :\Leftrightarrow i \leq j$ and let $a(j)$ be the multiplicity of the right hand side node $j$. Intuitively, the left hand side nodes represent elements and the right hand side nodes represent slots. For any subset $V \subseteq [n]$ of elements, the number of available slots is $\sum_{j \in \delta(V)} a(j) \geq \sum_{j=1}^{|V|} a(j) \geq |V|$ (denote $\delta(V) := \{j \mid \exists (i,j) \in E\}$), hence by Hall's theorem there exists a left perfect matching $M \subseteq E$ such that $\deg_M(j) \leq a(j)$ (with $\deg_M(j) := |\{i \mid (i,j) \in M\}|$). For every $(i,j) \in M$, we round up the element $i$ to the larger element $j$, then $\mathcal{S}'$ is a solution for the rounded instance. □